\documentstyle[preprint,epsfig,aps]{revtex}

\newcommand{\be}{\begin{equation}}
\newcommand{\ee}{\end{equation}}
\newcommand{\bea}{\begin{eqnarray}}
\newcommand{\eea}{\end{eqnarray}}
\newcommand{\half}{\mbox{$\textstyle \frac{1}{2}$}}
\newcommand{\ket}[1]{ | \, #1  \rangle}
\newcommand{\bra}[1]{ \langle #1 \,  |}
\newcommand{\proj}[1]{\ket{#1}\bra{#1}}
\newcommand{\abs}[1]{ | \, #1 \,  |}

\input{psfig.sty}

\begin{document}
\title{Optimal universal and state-dependent quantum cloning}
\author{Dagmar~Bru\ss $^{1,4}$, David~P.~DiVincenzo$^2$,
         Artur~Ekert$^1$, Christopher~A.~Fuchs$^3$,
         Chiara~Macchiavello$^{1,4}$, John~A.~Smolin$^2$}
\address{$^1$Clarendon Laboratory, University of Oxford, Parks Road,
      Oxford OX1 3PU, UK\newline
  $^2$IBM Research Division, T.J. Watson Research Center,
    Yorktown Heights, NY 10598, USA\newline
  $^3$Bridge Laboratory 12-33, California Institute of Technology,
Pasadena,
  CA 91125, USA\newline
  $^4$ISI, Villa Gualino, Viale Settimio Severo 65, 10133 Torino,
            Italy }
\date{Received \today}
\maketitle
\begin{abstract}
We establish the best possible approximation to a perfect quantum
cloning machine which produces two clones out of a single input.
We analyze both universal and state-dependent cloners.  The maximal
fidelity of cloning is shown to be 5/6 for universal cloners. It can
be achieved either by a special unitary evolution or by a novel
teleportation scheme. We construct the optimal state-dependent cloners
operating on any prescribed two non-orthogonal states, discuss their
fidelities and the use of auxiliary physical resources in the process
of cloning. The optimal universal cloners permit us to derive a new
upper bound on the quantum capacity of the depolarizing quantum
channel.
\end{abstract}
\narrowtext


\section{Introduction}
A $1\rightarrow 2$ {\it quantum cloner\/} is a quantum mechanical
machine that  transforms a system described by  some given
 pure state $\ket{\psi}$ together with some prescribed
 state into two systems, each with a
state as ``close'' as possible to the given one.  Specifically,
a quantum cloner for qubits is defined by an input qubit
$\ket{\psi}$, a blank qubit $\ket{0}$,
 an ancillary system $A$ in a state $\ket{X}$ (if necessary),
and a unitary transformation $U$ acting on  all three of these, such
that
\be
\ket{\psi}\ket{0}\ket{X}\;\longrightarrow\;\ket{\Psi}=
U\ket{\psi}\ket{0}\ket{X}\;,
\label{neweq}
\ee
and, after the interaction, the reduced density operators for
the two qubits are identical---i.e., if
$\rho_1={\mbox{Tr}}_{2,A}\big(|\Psi\rangle\langle\Psi|\big)$
and
$\rho_2={\mbox{Tr}}_{1,A}\big(|\Psi\rangle\langle\Psi|\big)$,
then $\rho_1=\rho_2$.  In general, ideal quantum cloners (i.e., ones
for which $\rho_1=\rho_2=\ket{\psi}\bra{\psi}$) do not exist: only
if $\ket{\psi}$ is assured to be drawn from a fixed orthogonal set
can such a quantum cloner be constructed
\cite{Wootters,Dieks,Yuen,Barnum}.
This situation, however, leads naturally to the question, ``How
close to ideal can a cloner be?''  This can be explored both as a
function of the sets from which the unknown state can be drawn and
as a function of various notions of ``closeness'' to ideality.  In
this paper, we explore two such sets and optimality criteria.

We define a {\em universal\/} $1\rightarrow 2$ quantum cloner as a
quantum machine that takes as an input one qubit in a completely
unknown quantum state $\ket{\psi}$ and generates at the output two
qubits such that each of them is in a state described by the reduced
density operator of the form
$\rho=\eta \proj{\psi} + (1-\eta) \frac{1}{2} {\bf 1}$.
The parameter $\eta$ describes the shrinking of the original
Bloch vector $\vec s$ corresponding to the density operator
$\proj{\psi}$, i.e. if $\proj{\psi} =\frac{1}{2}({\bf 1}+\vec
s\cdot\vec\sigma)$ then $\rho=\frac{1}{2}({\bf 1}+\eta\vec
s\cdot\vec\sigma)$, where ${\bf 1}$ is the 2$\times$2 identity matrix
and $\vec\sigma$ represents the set of Pauli matrices.  In this case,
we shall be interested in the best possible cloner with respect to
the criterion of maximal $\eta$---that is, maximal ``local'' fidelity
$F=\bra{\psi}\rho\ket{\psi}=\frac{1}{2}(1+\eta)$ between input and
output.  This case is important because it reveals the overall
effectiveness with which purely quantum information---embodied in
a completely unknown quantum state---can be copied.
\par
In some cases the original qubit may be prepared in a state which is
selected from a known ensemble of states. In such cases we can design
a {\em state-dependent\/} cloner which is optimal with respect to a
given ensemble; here we will consider ensembles composed of
only two non-orthogonal quantum states
$\ket{a}$ and $\ket{b}$.  Here the criterion of optimality is
that of optimizing the ``global fidelity'' between input and
output, i.e.,
to make the state $\ket{\Psi}$ given in Eq.~(\ref{neweq}) have
the largest inner product possible with $\ket{a}\ket{a}$
or $\ket{b}\ket{b}$, depending upon the input state.
This case is of some importance, for instance, because of the way
it compares and contrasts to optimal eavesdropping schemes on
two-state quantum cryptographic protocols \cite{bennett}.

The paper is organized as follows. In Section \ref{uqc} we discuss the
performance of a universal quantum cloner, analyzing the role of the
symmetry and isotropy conditions imposed on the system.  The cloning
transformation with the optimal ``local'' fidelity is derived by a
{\em constructive} proof and is shown to coincide (modulo some phase
factors) with the cloning machine proposed by Bu\v{z}ek and Hillery
\cite{vlado}. We then demonstrate in Section II B that universal
quantum cloners can also be implemented via quantum-state
teleportation \cite{B6}.  This method results in the creation of
two imperfect clones at two different locations by a combination of a
shared three particle entanglement and public broadcasting. In Section
\ref{sec:noan} we relax the universality requirement and study
state-dependent cloners. We derive the optimal cloning transformations
with respect to two-state input ensembles. We also comment on
the role of state-dependent cloners in quantum cryptography and
show that the ``local'' and ``global'' fidelity criteria lead to
distinct notions of cloning in the state-dependent case.
Finally in Section IV, as an application of these results, we
relate the optimality of universal cloners to quantum channel
capacity.  All technical details of the
optimality proofs are included in Appendix A (universal cloners) and
Appendix B (state-dependent cloners).  Appendix C details the
calculations required to compare the optimal state-dependent cloner
to optimal eavesdropping in quantum cryptography \cite{fuchs1,fuchs2}.

Let us point out again that in this paper we restrict our discussion
only to $1\rightarrow 2$ cloners.  More general results will be
presented elsewhere.

\section{Universal Quantum Cloner}

In subsection \ref{uqc} we derive the unitary transformation
with optimum fidelity for a universal $1\rightarrow 2$  quantum
cloner. We then show in subsection \ref{tele} the possibility of
establishing
this cloning transformation via teleportation.

\subsection{Optimal universal quantum cloner}
\label{uqc}

In this subsection we find the optimum fidelity for a quantum cloner
which is defined as a unitary transformation acting on two initial
qubits (the one to be cloned in state $\ket{\psi}$ (or
$\rho_\psi=\ket{\psi}\bra{\psi}$) and the second one in a standard
state $\ket{0}$) and an auxiliary system, also referred to as ancilla.

We will impose the following conditions on a universal quantum cloner:

\begin{tabular}{lll}
I. & $\rho_1=\rho_2$ & Symmetry \\
II.a) &  $\vec s_1=\eta_\psi \vec s_\psi$  & Orientation Invariance of
              Bloch vector \\
II.b)  &
             $F=\mbox{Tr}(\rho_\psi \rho_1) = \mbox{const.}$
           & Isotropy
\end{tabular} \\
$\rho_1$ and $\rho_2$ represent the reduced density operators of the
two output
qubits:
\be
\rho_1={\mbox{Tr}}_{2,A}[|\Psi\rangle\langle\Psi|]
\ee
where $\ket{\Psi}$ is the global state at the output of the cloner and
the
partial trace is performed on the second copy and the ancilla's degrees
of
freedom, and analogously for $\rho_2$.
$F$ is the fidelity of the cloner.
\par
Let us comment on these three conditions:
The first condition demands that the reduced density matrices of the
two output states are the same.  This is what we mean by symmetric
cloning.
\par
The second condition requires that the Bloch vector of the original
state
$\psi$ does not change its direction but only its length: it
shrinks by a factor $\eta_\psi$, indicating that the clones are not
pure states, due to entanglement between themselves and the ancilla.
\par  The third condition requires that
the cloner treats every state in the same way, i.e. the fidelity and
thus
the reduction factor
$\eta$ does not depend on the input vector.
\par We will see in the following that conditions II.a) and
II.b) are not
independent: if conditions I. and II.a) are satisfied II.b) holds
automatically,
i.e. symmetry plus orientation invariance implies isotropy.
\par
On the other hand we notice that any transformation on a qubit (i.e. on
a Bloch vector) 
can be decomposed into a transversal (rotation) and a longitudinal
(rescaling) part. By demanding that the cloner treats all input states
in the same way the Bloch vector of the original qubit can only be
rescaled but not rotated, because a rotation has always two fixed
points
on the sphere (`hairy ball' theorem), so at least two states are
transformed in a ``special" way which contradicts the universality
requirement.
\par
Thus for a symmetric cloner the second and the third condition are
equivalent. This is the reason that we
called them II.a) and II.b).

We start from a general ansatz for the
unitary transformation $U$ performed by the cloner
and acting on the total Hilbert space
${\cal H}^T = {\cal H}^2\otimes {\cal H}^2\otimes {\cal H}^x$,
where $x$ is the dimension of the Hilbert space for the ancilla states:
\bea
U\, \ket{0}\ket{0}\ket{X} & = & a\ket{00}\ket{A}+b_1\ket{01}\ket{B_1}
     +b_2\ket{10}\ket{B_2}+ c\ket{11}\ket{C}
 \label{eq:00}\\
U\, \ket{1}\ket{0}\ket{X} & = & \tilde a\ket{11}
         \ket{\tilde A}+\tilde b_1\ket{10}\ket{\tilde B_1}
     +\tilde b_2\ket{01}\ket{\tilde B_2}
     + \tilde c\ket{00}\ket{\tilde C}.
  \label{eq:10}
\eea
Here
$\ket{X}$ denotes the initial state of the ancilla.
Capital letters $A,B_i,C,...$ refer to output ancilla states.
We have not specified  the dimension of the ancilla, and we have not
assumed
any orthogonality relation to hold between
$\ket{A},\ket{B_i},...$\, . The only condition
we are imposing on $\ket{A},\ket{B_i},...$ is that they are
normalized. In this way we do not restrict our argument leading to the
optimum cloner to a certain dimension of the Hilbert space of the
ancilla. From this general ansatz we can also draw  conclusions about
the existence of symmetric and isotropic
quantum cloning without ancilla
which are discussed in Section \ref{sub:n1}.
\par Due to the unitarity of the cloning transformation, the
coefficients
$a,b_i,c,...$, which are in general complex, must satisfy the
normalization
conditions
\bea
\abs{a}^2+\abs{b_1}^2+\abs{b_2}^2+\abs{c}^2& = & 1 \nonumber \\
\abs{\tilde a}^2+\abs{\tilde b_1}^2+\abs{\tilde b_2}^2+
        \abs{\tilde c}^2& = & 1 \
\label{eq:norm}
\eea
and the orthogonality condition
\be
a^*\tilde c \bra{A}\tilde C \rangle +
b_2^*\tilde b_1 \bra{B_2}\tilde B_1 \rangle +
b_1^*\tilde b_2 \bra{B_1}\tilde B_2 \rangle +
c^*\tilde a \bra{C}\tilde A \rangle = 0.
\label{eq:uni}
\ee

We now impose the constraints I and II to satisfy the symmetry
and the isotropy properties.
We  define the free phases for the coefficients as
$a = \abs{a} e^{i\delta_a}$,
$\tilde a = \abs{\tilde a} e^{i\delta_{\tilde a}}$  and analogously for
the
other coefficients.
>From imposing the symmetry condition we find that our
ansatz (\ref{eq:00}), (\ref{eq:10}) has to fulfill the following
relations:
\bea
\abs{b_1}& = & \abs{b_2} \ ; \quad \abs{\tilde b_1}  =  \abs{\tilde
b_2}
\nonumber \\
| \langle B_1 \ket{\tilde B_2} | & = & | \langle B_2 \ket{\tilde B_1} |
         \ ; \quad
| \langle B_1 \ket{\tilde B_1} |  =  | \langle B_2 \ket{\tilde B_2} |
\label{eq:symfol}
\eea
and
\be
a b_1^* \langle B_1 \ket{A} + c^* b_2 \langle C \ket{B_2} =
a b_2^* \langle B_2 \ket{A} + c^* b_1 \langle C \ket{B_1}
\label{eq7}
\ee
and the same as Eq.~(\ref{eq7})
for the tilded coefficients and ancilla states. Moreover,
\bea
   \tilde b_1^* a\langle \tilde B_1 \ket{A} +
 \tilde a^* b_1\langle \tilde A \ket{B_1}
 & = &
  \tilde b_2^* a\langle \tilde B_2 \ket{A} +
 \tilde a^* b_2\langle \tilde A \ket{B_2}   \nonumber \\
   b_1^* \tilde c \langle B_1 \ket{\tilde C} +
   c^* \tilde b_1 \langle C \ket{\tilde B_1}
 & = &
   b_2^* \tilde c \langle B_2 \ket{\tilde C} +
   c^* \tilde b_2 \langle C \ket{\tilde B_2}  \ .
\eea
We will call $\abs{b_1} =  \abs{b_2} =  \abs{b}$ and
$|\, \tilde b_1\, |  =  |\, \tilde b_2\, |=  |\, \tilde b\, |$ from now
on.
\par
Let us  now look into the constraints following from imposing condition
II.a).
Orientation invariance of the Bloch vector
$\vec s$ means that all its components
shrink by the same ratio $\eta_\psi$:
\be
\frac{{s_1}_x}{{s_\psi}_x} = \frac{{s_1}_y}{{s_\psi}_y} =
            \frac{{s_1}_z}{{s_\psi}_z} = \eta_\psi.
\label{eq:condII}
\ee
Using the unitary transformation (\ref{eq:00}),(\ref{eq:10}) and
imposing
condition (\ref{eq:condII}) we find
the following constraints:
\begin{itemize}
\item[(i)]$\abs{a}^2-\abs{c}^2=\abs{\tilde a}^2-\abs{\tilde c}^2$
\item[(ii)]$\abs{a}^2-\abs{c}^2=\mbox{Re}
        \left[ \tilde b_1^* a \bra{\tilde B_1}A\rangle
            +\tilde a^* b_1 \bra{\tilde A}B_1\rangle \right] $
\item[(iii)]Im$\left[   \tilde b_1^* a\bra{ \tilde B_1}A\rangle
            +\tilde a^*  b_1 \bra{\tilde A} B_1\rangle \right] = 0$
\item[(iv)]$b_1^* \tilde c \bra{ B_1}\tilde C\rangle
            +c^* \tilde b_1 \bra{C}\tilde B_1\rangle = 0$
\item[(v)]$b_2^* a\bra{ B_2}A\rangle
            + c^* b_1\bra{C} B_1\rangle= 0 $
\item[(vi)]$ \tilde b_2^* \tilde a\bra{ \tilde B_2}\tilde A\rangle
            + \tilde c^* \tilde b_1\bra{\tilde C} \tilde B_1\rangle=0$
\item[(vii)]$ \tilde c^*a \bra{ \tilde C}A\rangle
            -\tilde a^* c\bra{\tilde A} C\rangle = 0$
\item[] and  $\ \ (1 \leftrightarrow 2)$
\end{itemize}
Here the notation $(1 \leftrightarrow 2)$  indicates that
as a result of the symmetry condition
the same set of
constraints has to hold for exchange of the indices 1 and 2.
\par
Inserting constraints (i) and (vii) into
the explicit form for the ratio $\eta_\psi$
we
find easily that  $\eta_\psi$ is a constant,
i.e. independent of the input state.
Thus, as mentioned before, we find that conditions II.a) and II.b) are
not
independent: after imposing condition I, condition II.b) is
automatically
satisfied when II.a) holds. Therefore, any symmetric cloner which does
not
rotate the initial state is isotropic.

The explicit form of the reduction factor $\eta$ is
\be
\eta=|a|^2-|c|^2
\label{eq:eta}
\ee
which we want to maximize. The fidelity
\be
F= \mbox{Tr}(\rho_1 \ket{\psi}\bra{\psi}) = \half (1+\vec s_1\cdot\vec
s_\psi),
\ee
which for the symmetric isotropic cloner is related to the reduction
factor as
\be
F=\half(1+\eta),
\ee
is maximized as well.
\par
The maximization of the fidelity is carried out using the Lagrange
multiplier
method, which takes into account the constraints imposed on the cloning
transformation due to the unitarity, symmetry and isotropy conditions.
Here we have also required the
unitary transformation to be symmetric under exchange
$\ket{0}\leftrightarrow
\ket{1}$ which leads to $\abs{a}=\abs{\tilde a},\abs{b}=\abs{\tilde b},
\abs{c}=\abs{\tilde c}$. The explicit optimization procedure is
reported in
Appendix \ref{app2}.
\par
The idea is to use the Lagrange multiplier technique and some
knowledge about the coefficients $a$ and $b_i$ to find the best value
for $\abs{c}$. Then we use constraint (ii)
to find the optimum value of $\abs{a}$ which  gives us $\abs{b}$ via
the normalization condition.
\par The results are
\bea
\abs{c}& =& 0 \nonumber \\
\abs{a} & = & \sqrt{\frac{2}{3}} \nonumber \\
\abs{b} & = & \sqrt{\frac{1}{6}} \ .
\eea
Here $\abs{c} = 0$ can be  understood intuitively because
$ c $ is the coefficient
for the state $\ket{11}$ which is
maximally  remote from the ideal  output state $\ket{00}$
in Eq.~(\ref{eq:00}).
\par
Thus we find that the shrinking factor $\eta$ of the optimum
symmetric isotropic cloner is
\be
\eta = \frac{2}{3} \ ,
\ee
corresponding to the optimum cloning fidelity
\be
F=\frac{5}{6} \ .
\ee
\par
As shown in appendix \ref{app2}
the class of unitary transformations for the optimal symmetric and
isotropic
cloner
 is given by
\bea
U\, \ket{0}\ket{0}\ket{X} & = &
\sqrt{\frac{2}{3}}e^{i\delta_a}\ket{00}\ket{A}+\sqrt{\frac{1}{6}}
      e^{i\delta_{\tilde a}}(\ket{01}
     +\ket{10})\ket{A_\bot}
 \label{eq:00op}\\
U\, \ket{1}\ket{0}\ket{X} & = & \sqrt{\frac{2}{3}}e^{i\delta_{\tilde
a}}\ket{11}
         \ket{A_\bot}+
       \sqrt{\frac{1}{6}}e^{i\delta_a}(\ket{01}
                        +\ket{10})\ket{A}
  \label{eq:10op}
\eea
where $\bra{A}A_\bot\rangle =0$.
\par We can realize this transformation with two-dimensional ancilla
states,
e.g. $\ket{A}=\ket{0}$,  $\ket{A_\bot}=\ket{1}$,
or any other orientation of $\ket{A}$.
These  possibilities are different from each other with respect to the
reduced density matrix of the ancilla qubit.
If we  choose $\delta_a=
\delta_{\tilde a}=0$
and $\ket{A}=\ket{0}$ we arrive at
the cloning transformation proposed by Bu\v{z}ek and Hillery
\cite{vlado},
one example for the optimum symmetric and isotropic cloner.
\par As the requirements for the scalar products of the
ancilla states for the optimum cloner
 can be met
by using ancilla states of dimension two  there is no better cloner
using higher dimensional ancillas.
\par
We note that maximizing the global fidelity, defined by
$F_g=\mbox{Tr}[(\rho_\psi \otimes \rho_\psi)\rho_{1,2}]$
where $\rho_{1,2}$
denotes the total output density matrix, traced over the ancilla,
leads to the same transformations (\ref{eq:00op}) and (\ref{eq:10op}).

\subsection{Universal cloning by teleportation}
\label{tele}
So far we viewed the cloner as a machine which clones quantum states
at a given location. There are, however, certain scenarios, especially
in quantum communication and cryptography, where  cloning is
followed by further processing which may involve sending the two
clones to two different locations. In these scenarios one may benefit
from ``non-local" cloning which can be achieved via teleportation.

Suppose that a sender (Alice) is to transmit an imperfect copy of her
qubit state to two receivers (Bob and Charlie); the three parties
possess as a starting resource a particular entangled quantum state,
but otherwise only classical communication is permitted from Alice to
Bob and Charlie.  This situation is essentially the three-party
generalization of the well-known {\em teleportation}
protocol\cite{B6}, in which Alice can transmit any qubit state to Bob
perfectly, provided that they share an entangled singlet state
$\ket{\Psi^-}=\frac{1}{\sqrt{2}}(|01\rangle-|10\rangle)$.  In this
protocol, Alice first performs a joint measurement of the state to be
teleported $|\psi\rangle$ and her half of the singlet pair, the
measurement being performed in the Bell basis
\begin{equation}
\begin{array}{l}
\ket{\Psi^\pm}=\frac{1}{\sqrt{2}}(|01\rangle\pm|10\rangle)\\
\ket{\Phi^\pm}=\frac{1}{\sqrt{2}}(|00\rangle\pm|11\rangle).
\end{array}
\end{equation}
Alice then sends a two-bit message to Bob indicating which of the four
Bell states was measured.  Bob can reconstitute $|\psi\rangle$ exactly
from his half of the singlet if he performs the final action: if he
receives the message ``$\Psi^-$'', nothing; if ``$\Psi^+$'', rotate
his qubit by $\sigma_z$; if ``$\Phi^-$'', rotate by $\sigma_x$; if
``$\Phi^+$'', rotate by $\sigma_y$.

The same protocol, but applied to the particular three-particle
state
\begin{equation}
\ket{\Psi_{clone}}=\sqrt{\frac{2}{3}}|100\rangle-
\sqrt{\frac{1}{6}}|010\rangle-\sqrt{\frac{1}{6}}|001\rangle
\end{equation}
results precisely in a Bu\v{z}ek-Hillery cloning from Alice to Bob and
Charlie, provided that the results are averaged over the four possible
measurement outcomes.  (This averaging is not necessary in ordinary
teleportation; we will explain in a moment what happens if the
measurement outcomes are {\em not} averaged over in the present form
of teleportation.)  In $\ket{\Psi_{clone}}$, the first particle is
possessed by Alice, the second by Bob, and the third by Charlie.  (Of
course, the state is symmetric with respect to Bob and Charlie.)  The
cloning is achieved by classical transmission in the sense that Alice
need only broadcast the two-bit result of her Bell measurement to Bob
and Charlie, with which they perform the same final action as in
teleportation, in order for Bob and Charlie to possess
Bu\v{z}ek-Hillery clones of Alice's original qubit state.

It is informative to formulate our imperfect teleportation in the
language of quantum operations with which Nielsen and Caves have
analyzed ordinary teleportation\cite{NC}.  In this language the
transformation from Alice's input state $\rho_\psi$ and Bob's (or
Charlie's) output state $\rho_o^i$ conditional upon measurement outcome
$i$ (unnormalized) is specified by the superoperator
\begin{equation}
\rho_o^i=\sum_{j}A_{ij}\rho_\psi A^\dagger_{ij}.
\end{equation}
The output density operator $\rho_o$
taking all measurement outcomes into account
 just requires the sum over all outcomes $i$:
\begin{equation}
\rho_o=\sum_{ij}A_{ij}\rho_\psi A^\dagger_{ij}.
\end{equation}
Completeness requires
\begin{equation}
\sum_{ij}A^\dagger_{ij}A_{ij}=1,
\end{equation}
but if the sum is restricted to a particular $i$,
\begin{equation}
\sum_{j}A^\dagger_{ij}A_{ij}=E_i,
\end{equation}
where $E_i$ is the operator representing the measurement outcome $i$
in the positive-operator-valued theory of quantum measurement.

This representation may be related simply to the Bloch-vector picture.
If we write the conditional output density operator as $\rho_o^i=
\frac{\rm Pr}(i){2}({\bf1}+{\vec s}_o\cdot{\vec\sigma})$, where
${\rm Pr}(i)$ is the probability of measurement outcome $i$, then
\begin{equation}
{\rm Pr}(i)=\half{\rm Tr}E_i+\half\sum_\alpha s_{\psi\alpha}{\rm
Tr}(E_i
\sigma_\alpha)
\end{equation}
and
\begin{equation}
{\rm Pr}(i)s_{o\beta}=
\half{\rm Tr}(\sum_jA_{ij}A_{ij}^\dagger\sigma_\beta)+
\half\sum_\alpha s_{\psi\alpha}{\rm Tr}(\sum_jA_{ij}\sigma_\alpha
A_{ij}^
\dagger\sigma_\beta).
\end{equation}

A straightforward calculation shows that for our imperfect
teleportation,
the $\Phi^+$ and $\Phi^-$ measurement outcomes are indistinguishable
(i.e., are described by the same $A$ operators, and therefore have the
same probability of occurrence and leave the output qubit in the
identical state).  This is also true of the $\Psi^+$ and $\Psi^-$
outcomes.  But the $\Phi$ and $\Psi$ measurements are distinct.  This
is by contrast to perfect teleportation in which all four measurement
outcomes lead to identical operations (just the trivial noiseless
identity operator, in fact).  For our case we find
\begin{equation}
\begin{array}{ll}
A_{\Phi,1}=
\sqrt{\frac{2}{3}}\left(\begin{array}{lr}\half&0\\0&
1\end{array}\right)&
A_{\Phi,2}=
\sqrt{\frac{1}{6}}\left(\begin{array}{lr}  0&0\\1&
0\end{array}\right)\\
\ &\ \\
A_{\Psi,1}=
\sqrt{\frac{2}{3}}\left(\begin{array}{lr}
1&0\\0&\half\end{array}\right)&
A_{\Psi,2}=
\sqrt{\frac{1}{6}}\left(\begin{array}{lr}  0&1\\0&  0\end{array}\right)
\label{aops}
\end{array}
\end{equation}
and
\begin{equation}
\begin{array}{ll}
E_\Phi=\left(\begin{array}{lr}1/3&0\\0&2/3\end{array}\right)&
E_\Psi=\left(\begin{array}{lr}2/3&0\\0&1/3\end{array}\right).\label{eops}
\end{array}
\end{equation}
We note from Eq.~(\ref{aops}) that the teleportation operation, keeping
only the cases where the measurement outcome is a $\Phi$, resembles in
some ways a ``decay channel'', in which the state is damped
towards the $|1\rangle$-state fixed point.  The cases where the
measurement
outcome is $\Psi$
behave identically except with $|0\rangle$ and $|1\rangle$
interchanged (from which the isotropy of the measurement-averaged
operation
emerges).  However, it is incorrect to say that the total operation is
obtained by selecting at random between the ``$\Phi$ channel'' and the
``$\Psi$ channel'' (although this is the way that \cite{gisin} creates
several interesting cloning transformations), because the $E_i$
operators are not proportional to the identity as they are in perfect
teleportation.  Unlike in the randomly-selected channel, the
probability of the measurement outcome depends on the input state; we
find directly from Eq.~(\ref{eops}) that
\begin{equation}
\begin{array}{ll}
{\rm
Pr}(\Phi)=1/3\langle0|\rho_\psi|0\rangle+2/3\langle1|\rho_\psi|1\rangle,
&
{\rm
Pr}(\Psi)=2/3\langle0|\rho_\psi|0\rangle+1/3\langle1|\rho_\psi|1\rangle.

\end{array}
\end{equation}

Finally we note that since the Bell measurements occupy a Hilbert
space of at least two qubits\cite{Brassard}, an open question is
raised of whether good 1-to-$N$ cloning can be achieved by
teleportation through an $N+1$-particle entangled state.  In the
simplest generalization of the above protocol the extra Hilbert space
size would still be two qubits (since there would still just be one
Bell measurement); but the optimal 1-to-$N$ cloner appears to require
an ancilla with $O(N)$ qubits\cite{thegang}; as V. Bu\v{z}ek has
pointed out, this may well mean that this teleportation approach to
cloning may not generalize to other cloning problems.

\section{State-dependent quantum cloners}
\label{sec:noan}

Let us start this section with showing  that in order to satisfy the
isotropy requirement
an ancilla system must be necessarily involved in the cloning
transformation.
This is proved in subsection \ref{sub:n1}.
In subsection \ref{sub:n2} we drop the isotropy condition and
investigate
the case of
a symmetric state dependent cloner in absence of ancilla. We will show
that
if we have some {\it a priori} knowledge about the input states the
cloner can
perform much better than the optimal universal one.

\subsection{Quantum cloner without ancilla}
\label{sub:n1}

>From the general ansatz for the unitary transformation with an ancilla
of arbitrary dimension we can draw  conclusions about a quantum
cloner without ancilla by replacing all states $\ket{A},\ket{B_i},...$
on the RHS of (\ref{eq:00}) and (\ref{eq:10}) with the factor 1.
\par
If we attempt to realize
 a symmetric and isotropic cloner
we need to be able to fulfill
 the constraints (i) to (vii) where all scalar products of auxiliary
states
have to be replaced by 1. We will show that this is not possible.
\par Here we only write down those four constraints which we need for
our
argument:
\begin{itemize}
\item[(i)]$\abs{a}^2-\abs{c}^2=\abs{\tilde a}^2-\abs{\tilde c}^2$
\item[(ii)]$\abs{a}^2-\abs{c}^2=\mbox{Re}\left[ \tilde b_1^* a
            +\tilde a^* b_1 \right] $
\item[(v)]$b_2^* a + c^* b_1= 0 $
\item[(vi)]$\tilde b_2^* \tilde a +\tilde  c^* \tilde b_1= 0$
\end{itemize}
Remember that from (\ref{eq:symfol}) we have
$\abs{b_1}=\abs{b_2}=\abs{b}$
and $\abs{\tilde b_1}=\abs{\tilde b_2}=\abs{\tilde b}$.
In order to fulfill constraints (v) and (vi) where both real and
imaginary part of the given sum have to vanish there are only these
possibilities (for any choice of phases $\delta_{a},\delta_{b_i},...$):
\be
\mbox{(v)} \leadsto \abs{b}=0 \ \ \ \mbox{or}\ \ \ \abs{a}=\abs{c}
\ee
and
\be
\mbox{(vi)} \leadsto \abs{\tilde b}=0 \ \ \ \mbox{or}\ \ \
                      \abs{\tilde a}= \abs{\tilde c} \ .
\ee
There are four possible combinations of these constraints:
\begin{itemize}
\item $\abs{a}= \abs{c}$ and $\abs{\tilde a}= \abs{\tilde c}$
\item $\abs{a}= \abs{c}$ and $\abs{\tilde b}=0$
\item $\abs{b}=0$ and $\abs{\tilde a}= \abs{\tilde c}$
\item $\abs{b}=0$ and $\abs{\tilde b}=0$
\end{itemize}
For the first three possibilities we find immediately from (i) and
equation (\ref{eq:eta}) that $\eta=0$, the trivial solution.
 For the last possibility we only need
a glance at constraint (ii) to find  $\eta=0$
as well.
\par
We thus conclude that it is impossible to build a symmetric isotropic
quantum cloner without ancilla.

\subsection{Optimal state-dependent cloner}
\label{sub:n2}
In this subsection we answer the following question: given
two possible input states $\ket{a}$ and $\ket{b}$, where
in general $\bra{a} b\rangle \ne 0$, what is the optimal
quantum cloner with respect to a ``global fidelity'' criterion?
We suppose that the input qubit is prepared with the same
probability either in state $\ket{a}$ or  $\ket{b}$ and
optimize the transformation as a function of their scalar
product.  The resulting optimal transformation will be
therefore state-dependent.

Two pure non-orthogonal states in a two-dimensional Hilbert
space can be parameterized as follows:
\begin{eqnarray}
\ket{a}&=&\cos \theta \ket{0}+\sin\theta \ket{1}  \nonumber \\
\ket{b}&=&\sin \theta \ket{0}+\cos\theta \ket{1},
\label{ketab}
\end{eqnarray}
where $\{\ket{0},\ket{1}\}$ represents an orthonormal basis  and
$\theta\in[0,\pi/4]$. The set of the two input states can equivalently
be specified by means of their scalar product $S=\langle a|b\rangle=
\sin 2\theta$.

Let us consider a unitary operator $U$ acting on
${\cal H}^T={\cal H}^2\otimes{\cal H}^2$  and define
the final states $\ket{\alpha}$ and $\ket{\beta}$ as
\begin{eqnarray}
\ket{\alpha}&=&U \ket{a}\ket{0}\\
\ket{\beta}&=&U \ket{b}\ket{0}.
\end{eqnarray}
Unitarity gives the following constraint on the scalar product of the
final states:
\begin{equation}
\langle\alpha|\beta\rangle=\langle a|b\rangle=\sin 2\theta\equiv S.
\label{prodscal}
\end{equation}

As a criterion for optimality of the state-dependent cloner, we
take the transformation that maximizes the global fidelity
$F_g$ of both final states $\ket{\alpha}$ and $\ket{\beta}$ with
respect to the perfect cloned states
$\ket{aa}\equiv\ket{a}\otimes\ket{a}$ and
$\ket{bb}\equiv\ket{b}\otimes\ket{b}$.  The global fidelity
is defined formally as
\begin{equation}
F_g =\frac{1}{2}\left(|\langle \alpha|aa\rangle|^2
+|\langle\beta|bb\rangle|^2\right).
\label{fidgen}
\end{equation}

We show in Appendix \ref{app4} that the above fidelity is maximized
when
the states $\ket{\alpha}$ and $\ket{\beta}$ lie in the two-dimensional
space
${\cal H}_{aa,bb}$ which is spanned by vectors $\{\ket{aa},
\ket{bb}\}$.

Let us now maximize explicitly the value of the global fidelity
(\ref{fidgen}). We can think about it in a geometrical way and define
$\phi$, $\delta$ and $\gamma$ as the ``angles'' between vectors
$\ket{aa}$ and
$\ket{bb}$, $\ket{aa}$ and $\ket{\alpha}$, $\ket{\alpha}$ and
$\ket{\beta}$ respectively. The global
fidelity (\ref{fidgen}) then takes the form
\begin{equation}
F_g=\frac{1}{2}\left(\cos^2\delta+\cos^2(\phi-\gamma-\delta)\right)
\label{fidgen2}
\end{equation}
and is thus maximized when the angle
between $\ket{aa}$ and $\ket{\alpha}$ is equal to the angle between
$\ket{bb}$ and $\ket{\beta}$, i.e. $\delta=\frac{1}{2}(\phi-\gamma)$.
The optimal situation thus corresponds to the maximal symmetry
in the disposition of the vectors.

As we know that $\phi=\arccos(\sin^2 2\theta)$ and from Eq.
(\ref{prodscal}) $\gamma=\arccos(\sin 2\theta)$, after little algebra
we can write the optimal global fidelity as
\begin{equation}
 F_{g,opt} =\frac{1}{4}\left(\sqrt{1+\sin^2 2\theta}\sqrt{1+\sin
2\theta}
+\cos2\theta \sqrt{1-\sin 2\theta}\right)^2.
\label{fidgenmax}
\end{equation}
The corresponding unitary transformation $U$ on the basis states
$\ket{00}$
and $\ket{10}$
of the initial subspace of the four-dimensional Hilbert space of the
two
qubits is given by
\begin{eqnarray}
U\ket{00}&=&a\ket{00}+b(\ket{01}+\ket{10})+c\ket{11}
\label{u-basis0} \\
U\ket{10}&=&c\ket{00}+b(\ket{01}+\ket{10})+a\ket{11}
\label{u-basis1}
\end{eqnarray}
where
\begin{eqnarray}
a&=&
{1\over{\cos 2\theta}}\left[\cos\theta(P+Q\cos 2\theta)
-\sin\theta(P-Q\cos 2\theta)\right]\\
b&=&{1\over{\cos 2\theta}}P\sin 2\theta(\cos\theta-\sin\theta)\\
c&=&{1\over{\cos 2\theta}}\left[
\cos\theta(P-Q\cos 2\theta)-\sin\theta(P+Q\cos 2\theta)\right]\;.
\label{A-B-C}
\end{eqnarray}
with
\begin{eqnarray}
P&=&\frac{1}{2}{{\sqrt{1+\sin 2\theta}}\over{\sqrt{1+\sin^2
2\theta}}}\\
Q&=&\frac{1}{2}{{\sqrt{1-\sin 2\theta}}\over{\cos 2\theta}}.
\label{coeff}
\end{eqnarray}
The transformation for
$\ket{\alpha}$ and $\ket{\beta}$ can be readily derived from Eqs.
(\ref{u-basis0}) and (\ref{u-basis1}). We can easily see that the
transformation
is symmetric, i.e.  $\rho_{\alpha,1}=\rho_{\alpha,2}=\rho_\alpha$ for
 input state $\ket{a}$ and similarly for $\ket{b}$.
\par
\begin{figure}[hbt]
\setlength{\unitlength}{1pt}
\begin{picture}(500,300)
\
\epsfysize=8cm
\epsffile[72 230 540 560]{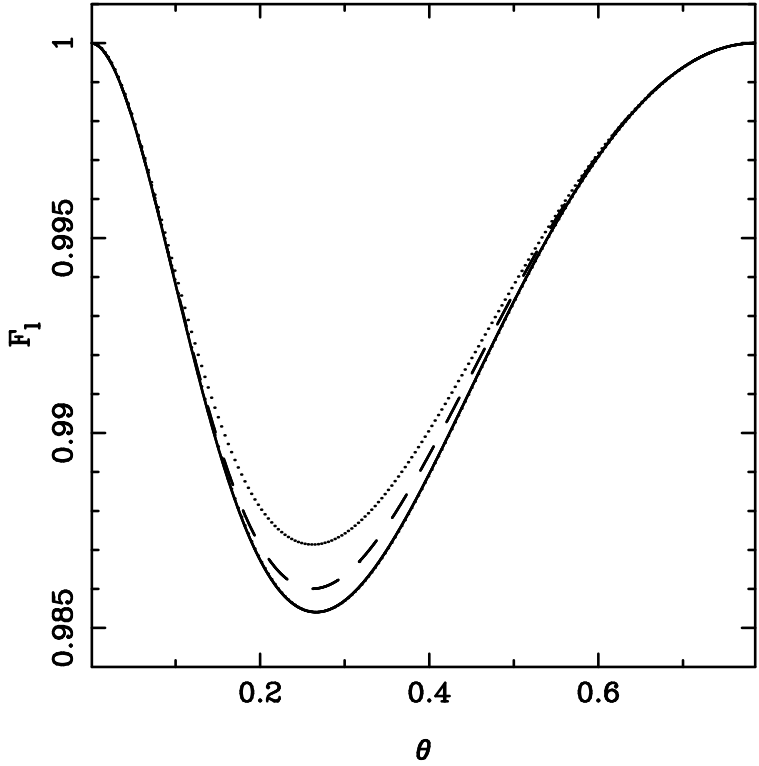}
\vspace{-0.2cm}
\end{picture}
\vspace{0.5cm}
\caption[]
        {\small Local fidelity of the state-dependent cloner as function
of $\theta$: the solid line results from  maximization of the global fidelity,
see eq. (\ref{uhlfid}), the dashed line corresponds to the local fidelity in 
the optimal
eavesdropping scheme, given in eq.
(\ref{Herbert}), and
the dotted line is the optimal local fidelity,
see eq. (\ref{Yehuda}).}
\label{f:uhlfid}
\end{figure}

In order to compare the performance of the state-dependent cloner with
the universal one we calculate the ``local'' fidelity $F_l$ of each of
the
output copies with respect to the input one,
generally defined as
\be
F_l={\mbox{Tr}}[\rho_{\alpha}|a\rangle\langle a|] \ \ .
\ee
For the above transformation we find
\begin{eqnarray}
F_{l,1}
&=&\frac{1}{2}\left[1+{{\cos^2 2\theta}\over{\sqrt{1+\sin^2 2\theta}}}
+{{\sin^2 2\theta(1+\sin 2\theta)}\over{1+\sin^2 2\theta}}\right].
\nonumber \\
&=&
\frac{1}{2}\left[1+\frac{1-S^2}{\sqrt{1+S^2}}+\frac{S^2(1+S)}{1+S^2}\right]
\label{uhlfid}
\end{eqnarray}
Due to  the symmetry of the problem the same expression (\ref{uhlfid})
is
obtained for the
fidelity of $\rho_{\beta}$ 
and it is plotted in Fig. \ref{f:uhlfid}. As we can see, the
fidelity takes surprisingly high values in the whole
range of $\theta$, well above the optimal value 5/6 of the universal
cloner.

\begin{figure}[hbt]
\setlength{\unitlength}{1pt}
\begin{picture}(500,300)
\
\epsfysize=8cm
\epsffile[72 230 540 560]{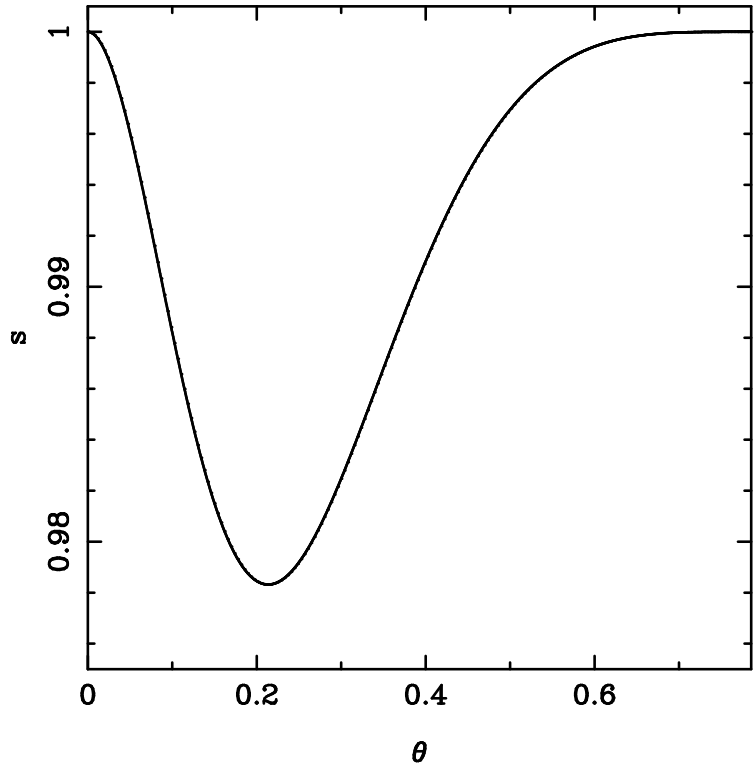}
\vspace{-0.2cm}
\end{picture}
\vspace{0.5cm}
\caption[]
        {\small Modulus of Bloch vector $\vec s$  for optimal
state-dependent
           cloner, see eq. (\ref{s_vector}).}
\label{f:s_vect}
\end{figure}

Let us now examine the degree of entanglement that our
``quasi-cloning'' transformation has introduced in the system.
An estimation of the degree of purity of the state is
given by the modulus of the $\vec s$ vector in the Bloch sphere:
the modulus is maximized to unit when the state is pure. In the case
under consideration it takes the form
\begin{equation}
|{\vec{s}}|=\sqrt{{{\sin^2 2\theta(1+\sin
      2\theta)^2}\over{(1+\sin^2 2\theta)^2}}+ {{\cos^2 2\theta}
\over{1+\sin^2 2\theta}}}
\label{s_vector}
\end{equation}
and is plotted in Fig. \ref{f:s_vect}. As we can see, the top of the
vector $\vec s$ is always very close to the surface of the Bloch sphere
for
any value of $\theta$ and the degree of purity of the output state
 is therefore always fairly high.  Notice that the length of the
Bloch vector is always much bigger than the value 2/3 of the
optimal universal cloner.
\par We also point out that in this case the Bloch vector is not only
shrunk
but also rotated by a state-dependent angle $\vartheta$, given by
\be
\vartheta = \arccos \left[ \frac{1}{\abs{\vec s}} \frac{\cos 2\theta}
      {\sqrt{1+\sin^2 2\theta}} \right] - 2\theta \ .
\ee

Perhaps the most important practical use for state-dependent cloners is
in
the eavesdropping on some quantum cryptographic systems. For example,
if
the quantum key distribution protocol is based on two non-orthogonal
states
\cite{bennett}, the optimal state dependent cloner can clone the qubit
in
transit between a sender and a receiver. The original qubit can then be
re-sent to the receiver and the clone can stay with an eavesdropper who
by
measuring it can obtain some information about the bit value encoded in
the
original.  The eavesdropper may consider storing the clone and delaying
the
actual measurement until any further public communication between
the sender and the receiver takes place.  This eavesdropping strategy,
for
instance, has been discussed recently in Ref.~\cite{gisin}.

It should be noted, however, that eavesdropping via a direct
cloning attempt is not the most advisable course of action for the
eavesdropper if she wishes to be the most surreptitious.  For that
task,
the eavesdropper's main concern is not in copying the quantum
information---as embodied in the two {\em non-orthogonal\/} quantum
states---but rather in optimizing the tradeoff between the classical
information made available to her versus the disturbance inflicted upon
the
original qubit~\cite{fuchs1,fuchs2}.  The optimal solution to that
problem leads to a one parameter class of unitary
interactions, the parameter being the degree of disturbance.  It turns
out that, regardless of the value of the parameter, the optimal unitary
interaction there never matches that given in
Eqs.~(\ref{u-basis0})--(\ref{coeff}).

Indeed this can be seen in a direct manner.  The optimal eavesdropping
strategy is quite similar to the scenario described above.  The
eavesdropper uses a probe system to interact with the in-transit qubit
and
then later performs a measurement on it (after all public discussion
has
ceased).  Although it is not assumed, it turns out to be sufficient to
take the probe system itself to be a single qubit \cite{fuchs1,fuchs2}.
In general the final state of the probe will not be the same as that of
the receiver's qubit:  for instance, if the eavesdropper's available
information is adjusted to vanish, then her probe will be left in its
original state, which is completely independent of the sender's qubit's
state.  Nevertheless, in Appendix C it is shown that when the
disturbance
is adjusted so that the statistical distinguishability between the
states
of the eavesdropper's probe is identical to that of the final states of
the
receiver's qubit, then the optimal eavesdropping scheme is actually a
quantum cloner.  In that case, the ``local'' fidelity between input
and output works out to be
\begin{equation}
F_{l,2}=\frac{1}{2}+\frac{\sqrt{2}}{4}\sqrt{
(1-2S^2+2S^3+S^4)+(1-S^2)\sqrt{(1+S)(1-S+3S^2+S^3)}}\;.
\label{Herbert}
\end{equation}
The difference between this fidelity and $F_{l,1}$ in
Eq.~(\ref{uhlfid})
is only slight---they differ at most by $0.000651$ when
$S=0.579924$, see Fig.~\ref{f:uhlfid}---but this is enough 
to show that optimal
cloning and optimal eavesdropping are two different tasks.

Similar results can be obtained for the four states in the BB84
quantum cryptographic protocol~\cite{BB84}.  Modifying the
optimal eavesdropping scheme for that protocol in Ref.~\cite{FGGNP}
into a quantum cloning device as above gives a local fidelity of
0.854.  
Note that in the  scenario of BB84 we can restrict the input
of a cloning machine, therefore
one would not want to use the universal
Bu\v{z}ek-Hillery cloner for the task of eavesdropping on BB84.

A more intriguing point, however, can be gleaned from noting that
actually
for all $S$, $F_{l,2}\ge F_{l,1}$.  This implies that the optimal
``global''
quantum cloner is not optimized with respect to the ``local'' fidelity
criterion:  in the state-dependent case, the two criteria differ.  In
fact,
the state-dependent cloner derived from optimal eavesdropping is still
not
the best with respect to the ``local'' fidelity criterion.  For
instance,
in Appendix C it is shown that there is a still better state-dependent
cloner for this criterion; it gives rise to a local fidelity given by
\begin{equation}
F_{l,3}=\frac{1}{2}+\frac{\sqrt{2}}{32S}(1+S)\Big(3-3S+\sqrt{1-2S+9S^2}\Big)
\sqrt{-1+2S+3S^2+(1-S)\sqrt{1-2S+9S^2}}.
\label{Yehuda}
\end{equation}
Again, the difference between $F_{l,3}$ and $F_{l,2}$
is not large---the largest
difference $0.001134$ is attained when $S=1/2$, see 
Fig.~\ref{f:uhlfid}---but it is enough to
show that there are better cloners out there with respect to the local
fidelity criterion.
We have verified that $F_{l,3}$ is indeed the optimal {\em local}
fidelity for a state-dependent cloner as defined in (\ref{u-basis0})
and (\ref{u-basis1})
but refrain from presenting the tedious calculations here.
Ultimately, the disparity between Eqs.~(\ref{uhlfid}), (\ref{Herbert})
and (\ref{Yehuda}) only points out the subtlety of the concept of
``copying'' quantum information: given that it cannot be done ideally,
there is no single sense in which it can be done in the best possible
way.

Finally, let us note that in this subsection we have always considered
qubits for the purpose of
illustration, but we stress that the results hold for an arbitrary
dimension of the input states. In this case we can rephrase our
arguments
in terms of the two-dimensional subspace spanned by the two input
states
and choose the same parametrization as given in Eq.~(\ref{ketab}) for
the
input states in such subspace. We can then derive the same conclusions
as above.

\section{Application to Quantum Capacity}
\label{conc}

The optimal universal cloners, e.g. the Bu\v{z}ek-Hillery cloner,
permit
us to establish a new upper bound on the quantum capacity of a
depolarizing qubit channel.
A simple $(1-\eta)$-depolarizing channel transmits a quantum state whose
Bloch vector
is shrunk by $\eta$, as above.  The quantum capacity $Q(\eta)$ is the
maximum rate at which $k$ qubits can be coded into $n$ qubits in such a
way
that the $k$ qubits can be recovered with high fidelity by the
receiver, in
the limit of $k$ and $n$ going to infinity.  We can show that
\begin{eqnarray}
&&Q=0,\;\eta \le {2\over 3}\nonumber\\
&&Q(\eta)\le 1-H_2\left({3\over 4} \eta + {1\over 4}\right),\;\eta>{2\over 3}
\label{thebound}
\end{eqnarray}
where $H_2(x)=-x \log_2 x - (1-x) \log_2(1-x)$ is the binary entropy
function.

The second part of Eq. (\ref{thebound}) is proved in \cite{Rains,Plenio}.
The proof of $Q=0$ for $\eta={2\over 3}$
follows from the universal cloning results above. Suppose the opposite,
$Q(\eta=\frac{2}{3})>0$; Sec. IV of \cite{B4} shows that
this cannot be so: Consider the Bu\v{z}ek-Hillery cloner inserted into
a three party Alice-Bob-Charlie communications protocol discussed
above in Sec. \ref{tele}.  If Bob and Charlie were oblivious to
each other's existence, they could both, by experiments conducted in
concert with Alice, establish that the Alice-Bob channel and the
Alice-Charlie channel are
both simple depolarizing channels with depolarization fractions
$\eta=\frac{2}{3}$.  If $Q(\frac{2}{3})>0$ this would mean that Alice
could, with suitable encoding, transmit a state to Bob and Charlie,
both of whom could successfully decode it and obtain a high-fidelity
copy of it.  But this violates the no-cloning theorem for quantum
states \cite{Wootters}; thus, it must be so that $Q(\frac{2}{3})=0$.
$Q=0$ for $\eta < {2 \over 3}$ follows from the non-decreasing
(as a function
of $\eta$) property of $Q$:  If a lower $\eta$ gave a higher $Q$ then
Alice could add noise herself to the signal thereby turning a high-$\eta$
channel into the supposedly better lower-$\eta$ channel.

The bound given in (\ref{thebound}) is discontinuous at $\eta={2 \over
3}$.  If we made the seemingly natural assumption that $Q$ is a
continuous function of $\eta$, as is the channel capacity in the
classical setting, then we can apply the methods introduced in
\cite{B4} to show that $Q\le 3 \eta -2$ for $\eta > {2\over 3}$.  This
would improve on (\ref{thebound}) for a range of $\eta$s near ${2\over
3}$.  Unfortunately, the continuity of $Q$ has proved suprisingly
difficult to establish rigorously; this has finally been established
\cite{BST97} for a particular channel, the quantum erasure channel.
The fact that the Bu\v{z}ek-Hillery cloner is proved to be optimal
shows that no stronger bound on $Q$ for the depolarizing channel can
be established by this reasoning, and in fact no upper bound with a
lower threshold is known, although there is also no evidence that the
capacity of Eq.~(\ref{thebound}) can be attained.  Thus this remains
one of the many open questions in quantum information theory.

\section*{Acknowledgements}
We thank Vladimir Bu\v{z}ek and Nicolas Gisin for discussions.
This work was supported in part by the European TMR Research Network
ERB-4061PL95-1412 and a NATO Collaborative Research Grant.
DPV and JAS thank the Army Research Office for support.
CAF was supported by a Lee A. DuBridge Fellowship and by DARPA through
the
Quantum Information and Computing (QUIC) Institute administered by the
US Army Research Office.
Part of this work was completed during the 1997 Elsag-Bailey -
I.S.I. Foundation research meeting on quantum computation.

\appendix
\section{Optimization for universal cloner}
\label{app2}

We want to maximize the function $\eta$ while the constraints (i) to
(vii)
as well as the  unitarity constraints (\ref{eq:norm}) and
(\ref{eq:uni}) are
fulfilled.
The  independent
variables are the  absolute values of the coefficients $a,b_i,...$,
their  phases,
the absolute values of
the scalar products of the ancilla states (2 of these are already fixed
via
the symmetry condition (\ref{eq:symfol})) and their phases
 which we notate as
\be
\bra{A}C\rangle = \abs{\bra{A}C\rangle}\cdot e^{i\delta_{AC}}
\ee
and accordingly for the other scalar products.
\par
We impose the natural symmetry requirement on the general ansatz that
the reduced density matrix of the two clones
  should not change under the exchange
$\ket{0}\leftrightarrow\ket{1}$,
i.e. the outcome should not depend on renaming the basis. This leads us
immediately to
\be
\abs{a}=\abs{\tilde a} \ ; \ \ \
\abs{b_i}=\abs{\tilde b_i} \ ; \ \ \
\abs{c}=\abs{\tilde c} \  \ \ \
\label{eq:notil}
\ee
and the following restrictions for the
scalar products of ancilla states from off-diagonal density matrix
elements:
\be
\abs{\bra{A}B_i\rangle}=\abs{\bra{\tilde A}\tilde B_i\rangle}; \ \ \
\abs{\bra{B_i}C\rangle}=\abs{\bra{\tilde B_i}\tilde C\rangle}; \ \ \
\abs{\bra{C}A\rangle}=\abs{\bra{\tilde C}\tilde A\rangle} \ \ \ .
\label{eq:notilsc}
\ee
We also find that the phases $\delta_{\tilde A \tilde B_i},
\delta_{\tilde B_i \tilde C},\delta_{\tilde C \tilde A}$ can be
expressed as functions of the phases
$\delta_{ A  B_i}, \delta_{B_i  C},\delta_{ C  A}$ and
$\delta_{a},\delta_{b_i},...$\, .
\par We are using the method of Lagrange multipliers
where we have to solve the system of equations
\bea
\frac{\partial \eta}{\partial \abs{a}} +\sum_{i=1}^{13} \lambda_i \cdot
\frac{\partial \varphi_i}{\partial \abs{a}} & = & 0 \nonumber \\
\frac{\partial \eta}{\partial \abs{b}} +\sum_{i=1}^{13} \lambda_i \cdot
\frac{\partial \varphi_i}{\partial \abs{b}} & = & 0 \nonumber \\
& & \nonumber \\
& ...  & \nonumber \\
& & \nonumber \\
\varphi_i & \equiv & 0 \ , \ \ \ \ \ \ i=1,...13
\eea
where
\be
\eta = 2\abs{a}^2+2\abs{b}^2-1 \ ,
\ee
 $\varphi_i$ denotes the  constraints, and the  Lagrange multipliers
are
$\lambda_i$.
The order of the  constraints which defines
the Lagrange multiplier indices in later equations is taken to be
\bea
\varphi_1& =& \abs{a}^2+2\abs{b}^2+\abs{c}^2-1 \nonumber \\
 \varphi_2& = &2\abs{a}^2+2\abs{b}^2-1-\mbox{Re}\left[ \tilde b_1^* a
            \bra{\tilde B_1}A\rangle
            +\tilde a^* b_1\bra{\tilde A}B_1\rangle \right] \nonumber
\\
 \varphi_3 &=&\mbox {Im}\left[   \tilde b_1^* a\bra{ \tilde
B_1}A\rangle
            +\tilde a^*  b_1 \bra{\tilde A} B_1\rangle \right]
\nonumber \\
 \varphi_4 &= & b_1^* \tilde c \bra{ B_1}\tilde C\rangle
            +c^* \tilde b_1 \bra{C}\tilde B_1\rangle  \nonumber \\
 \varphi_5& = &  b_2^* a\bra{ B_2}A\rangle
            + c^* b_1\bra{C} B_1\rangle  \nonumber \\
 \varphi_6& = &  \tilde b_2^* \tilde a\bra{ \tilde B_2}\tilde A\rangle
            + \tilde c^* \tilde b_1\bra{\tilde C} \tilde B_1\rangle
                                                 \nonumber \\
\varphi_7 &=& \tilde c^*a \bra{ \tilde C}A\rangle
            -\tilde a^* c\bra{\tilde A} C\rangle  \nonumber \\
\varphi_8 &=&a^*\tilde c \bra{A}\tilde C \rangle +
b_2^*\tilde b_1 \bra{B_2}\tilde B_1 \rangle  +
b_1^*\tilde b_2 \bra{B_1}\tilde B_2 \rangle  +
c^*\tilde a \bra{C}\tilde A \rangle  \nonumber \\
& \mbox{and} & \ \ \varphi_{9,10,11,12,13} \ = \ \varphi_{2,3,4,5,6}
                 \ \ \  \mbox{with} \ \ (1\leftrightarrow 2) \ .
\label{eq:lindex}
\eea
\par
In solving this system of equations we can use
some  knowledge
about the coefficients:
We know from constraint (ii) and
equation (\ref{eq:notil}) that both $\abs{a}$ and $\abs{b}$ can not
take the value 0, because otherwise $\eta=0$, the trivial solution.
\par
Taking the partial derivative with respect to $\abs{c}$ leads to
\bea
2 \lambda_1 \abs{c} & + &\lambda_4 \left[
               b_1^*e^{i\delta_{\tilde c}}\langle B_1\ket{\tilde C}+
               \tilde b_1e^{-i\delta_{c}}\langle C\ket{\tilde B_1}
\right]
               \nonumber \\
        & + &\lambda_{11} \left[
               b_2^*e^{i\delta_{\tilde c}}\langle B_2\ket{\tilde C}+
               \tilde b_2e^{-i\delta_{c}}\langle C\ket{\tilde B_2}
\right]
               \nonumber \\
        &+ &\lambda_5
                b_1e^{-i\delta_{c}}\langle C\ket{B_1} +
       \lambda_{12}
                b_2e^{-i\delta_{c}}\langle C\ket{B_2}
               \nonumber \\
      &+ &\lambda_6
               \tilde b_1e^{-i\delta_{\tilde c}}\langle\tilde C
         \ket{\tilde B_1} +
       \lambda_{13}
               \tilde  b_2e^{-i\delta_{\tilde c}}\langle \tilde C
       \ket{\tilde B_2}
               \nonumber \\
     & + & 2\lambda_8
               a^*e^{i\delta_{\tilde c}}\langle A
         \ket{\tilde C} = 0
\label{star}
\eea
where we have already eliminated $\langle \tilde A\ket{C}$ by
inserting $\varphi_7 \equiv 0$ into $\varphi_8$.  From the derivatives
with respect to $|\, \langle B_i \ket{\tilde C}\, |$,
$|\, \langle C \ket{\tilde B_i}\, |$,
$|\, \langle C \ket{B_i}\, |$,
$|\, \langle \tilde C \ket{\tilde B_i}\, |$
and $|\, \langle A \ket{\tilde C}\, |$
we arrive (after dividing through  phase factors) at
\be
\lambda_j \, \abs{b}\,  \abs{c} = 0 \hspace{1cm}
       \mbox {with} \hspace{0.5cm}j=4,5,6,11,12,13
\
\ee
and
\be
\lambda_8 \, \abs{a}\, \abs{c} = 0
\ .
\ee
After multiplying Eq.~(\ref{star}) with $\abs{c}$ we find
\be
\lambda_1\cdot \abs{c}^2 =0 \ \ .
\label{eq0}
\ee
In the same way we use the equations resulting from differentiating
with respect to $\abs{a}$, $|\, \langle \tilde B_i \ket{A}\, |$,
$|\, \langle \tilde A \ket{B_i}\, |$,
$|\, \langle B_i \ket{A}\, |$,
$|\, \langle \tilde B_i \ket{\tilde A}\, |$ and
$|\, \langle A \ket{\tilde C}\, |$ and get
\be
2 \abs{a}+\lambda_1 \abs{a} + 2 \lambda_2 \abs{a}
     + 2 \lambda_9 \abs{a}= 0 \ .
\ee
Multiplying this with $\abs{c}^2$ and using Eq.~(\ref{eq0}) we conclude
that,
since $\abs{a}\neq 0$,
  either $\lambda_2+\lambda_9 = -1$ or $\abs{c}=0$.
\par We will now show that $\lambda_2 +\lambda_9= -1$ corresponds to a
minimum
of $\eta$, i.e. $\eta=0$. From the derivatives with respect to
$|\, \langle \tilde {B_1} \ket{A}\,  |$ and
$|\, \langle \tilde {B_2} \ket{A}\,  |$ we find
after dividing through $\abs{a} \abs{b}$:
\bea
-\lambda_2  \cos(\delta_a -\delta_{\tilde b_1}+
         \delta_{\tilde B_1 A})
+\lambda_3  \sin(\delta_a -\delta_{\tilde b_1}+
         \delta_{\tilde B_1 A}) & =& 0
\label{aetz} \\
-\lambda_{9}  \cos(\delta_a -\delta_{\tilde b_2}+
         \delta_{\tilde B_2 A})
+\lambda_{10}  \sin(\delta_a -\delta_{\tilde b_2}+
         \delta_{\tilde B_2 A}) & =& 0
\label{lam910}
\eea
and from the derivatives with respect to $\delta_{\tilde B_1A}$
and $\delta_{\tilde B_2A}$:
\bea
\lambda_2  |\, \langle \tilde {B_1} \ket{A}\,  |
\sin(\delta_a -\delta_{\tilde b_1}+
         \delta_{\tilde B_1 A})
+\lambda_3  |\, \langle \tilde {B_1} \ket{A}\,  |
\cos(\delta_a -\delta_{\tilde b_1}+
         \delta_{\tilde B_1 A}) =0
\label{derdel} \\
\lambda_9  |\, \langle \tilde {B_2} \ket{A}\,  |
\sin(\delta_a -\delta_{\tilde b_2}+
         \delta_{\tilde B_2 A})
+\lambda_{10}  |\, \langle \tilde {B_2} \ket{A}\,  |
\cos(\delta_a -\delta_{\tilde b_2}+
         \delta_{\tilde B_2 A}) =0 \ .
\label{derdelb}
\eea
 If $\lambda_2 +\lambda_9= -1$ then
at least one of these two multipliers is not equal to zero. Let us
assume
that $\lambda_2\neq 0$.
We multiply Eq.~(\ref{aetz}) by $\cos(\delta_a -\delta_{\tilde b_1}+
\delta_{\tilde B_1 A})$, obtaining
\be
\lambda_3\cos(\delta_a -\delta_{\tilde b_1}+\delta_{\tilde B_1 A})
\sin(\delta_a -\delta_{\tilde b_1}+\delta_{\tilde B_1 A})=\lambda_2
\cos^2(\delta_a -\delta_{\tilde b_1}+\delta_{\tilde B_1
A}).\label{ddveq}
\ee
Substituting Eq.~(\ref{ddveq}) into Eq.~(\ref{derdel}) multiplied by
$\sin(\delta_a -\delta_{\tilde b_1}+\delta_{\tilde B_1 A})$ we obtain
$\lambda_2|\, \langle \tilde {B_1} \ket{A}\, |=0$, so that
\be
\langle \tilde {B_1} \ket{A} = 0 \hspace{0.4cm} \mbox{if}
\hspace{0.3cm}
      c\neq 0 \ .
\ee
The same reasoning in which tilded and untilded variables are
interchanged
leads to
\be
\langle \tilde {A} \ket{B_1} = 0 \hspace{0.4cm} \mbox{if}
\hspace{0.3cm}
      c\neq 0 \  \ .
\ee
Due to constraint (ii) this means $\eta=0$. \par
If the assumption $\lambda_2\neq 0$ does not hold then $\lambda_9\neq
0$
and the same line of arguments leads to
$\langle \tilde {B_2} \ket{A} = \langle \tilde {A} \ket{B_2} =0$
and also  $\eta=0$.
\par We have thus established
$\abs{c}=0$, and therefore $\eta=\abs{a}^2$.
\par We also notice that from $\varphi_5, \varphi_{12}$ and
$\varphi_6, \varphi_{13}$ we need
\be
\bra{B_i} A\rangle = \bra{\tilde B_i} \tilde A\rangle =0\ .
\ee
\par Now  $\eta$ is maximized by maximizing $\abs{a}$ which can be
easily
achieved using constraint (ii) and the normalization condition:
\bea
\mbox{(ii)}\leadsto \abs{a}^2 & = & \mbox{Re}\left[ \tilde b_1^* a
            \bra{\tilde B_1}A\rangle
            +\tilde a^* b_1\bra{\tilde A}B_1\rangle \right] \nonumber
\\
     & = & \abs{a}\cdot \abs{b} \cdot \mbox{Re}\left[
          e^{i(\delta_a- \delta_{\tilde b_1})}
            \bra{\tilde B_1}A\rangle
            +e^{i(\delta_{b_1}-\delta_{\tilde a})}
                        \bra{\tilde A}B_1\rangle \right]
    \nonumber \\
     & = & \abs{a}\cdot \abs{b} \cdot \xi
\eea
or
\be
\abs{a}^2  = 1-\frac{2}{2+\xi^2} \ .
\ee
So, $\abs{a}^2$ is maximized for the maximum value of $\xi^2$, which is
$\xi^2 = 2^2$. This leads  to
\be
\abs{a}  =  \sqrt{\frac{2}{3}}  \ ; \hspace{0.8cm}
\abs{b}  =  \sqrt{\frac{1}{6}} \ .
\ee
We can meet the maximum of $\xi$ by choosing
\be
\bra{\tilde A} B_i\rangle  =  1 \ ; \hspace{0.8cm}
\bra{\tilde B_i} A\rangle  =  1
\ee
and
\be
\delta_a  =  \delta_{\tilde b_i} \ ; \hspace{0.8cm}
 \delta_{\tilde a}  =   \delta_{b_i} \ .
\ee
Collecting our information about the
coefficients and scalar products the
class of optimal unitary transformations is given by
\bea
U\, \ket{0}\ket{0}\ket{X} & = &
\sqrt{\frac{2}{3}}e^{i\delta_a}\ket{00}\ket{A}+\sqrt{\frac{1}{6}}
      e^{i\delta_{\tilde a}}(\ket{01}
     +\ket{10})\ket{A_\bot}
 \label{eq:00opg}\\
U\, \ket{1}\ket{0}\ket{X} & = & \sqrt{\frac{2}{3}}e^{i\delta_{\tilde
a}}\ket{11}
         \ket{A_\bot}+
       \sqrt{\frac{1}{6}}e^{i\delta_a}(\ket{01}
                        +\ket{10})\ket{A}
  \label{eq:10opg}
\eea
where $\bra{A}A_\bot\rangle =0$.

\section{Optimization for state-dependent cloner}
\label{app4}
\par
Let us assume that $\ket{\alpha}$ and
$\ket{\beta}$ have some contribution which does not lie in
${\cal H}_{aa,bb}$. Then we can write explicitly the form of
$\ket{\alpha}$ and
$\ket{\beta}$
\begin{eqnarray}
\ket{\alpha} & = & a_0\ket{aa}+b_0\ket{bb}+ c_0\ket{C_0}
 \label{eq-u-a}\\
\ket{\beta} & = & a_1\ket{aa}+b_1\ket{bb}+ c_1\ket{C_1}
  \label{eq-u-b}
\end{eqnarray}
where vectors $\ket{C_0}$ and $\ket{C_1}$ are normalized and lie in the
subspace orthogonal to ${\cal H}_{aa,bb}$.
The unitarity of the transformation imposes the following constraints
\begin{eqnarray}
\varphi_1&=&\mbox{Re}[a_0^*a_1+b_0^*b_1+S^2(a_0^*b_1+b_0^*a_1)+
c_0^*c_1\langle C_0\ket{C_1}]-S=0
 \label{phi1}\\
\varphi_2&=&\mbox{Im}[a_0^*a_1+b_0^*b_1+S^2(a_0^*b_1+b_0^*a_1)+
c_0^*c_1\langle C_0\ket{C_1}]=0
  \label{phi2}\\
\varphi_3&=&|a_0|^2+|b_0|^2+2S^2\mbox{Re}[a_0^*b_0]+|c_0|^2-1=0
  \label{phi3}\\
\varphi_4&=&|a_1|^2+|b_1|^2+2S^2\mbox{Re}[a_1^*b_1]+|c_1|^2-1=0 \ ,
  \label{phi4}
\end{eqnarray}
where $S$ is defined in Eq.~(\ref{prodscal}).
\par
The global fidelity is given by
\begin{equation}
F_g=\frac{1}{2}\left(|a_0+b_0 S^2|^2+|b_1+a_1 S^2|^2\right).
\label{fid-ab}
\end{equation}
Inserting constraints $\varphi_3$ and $\varphi_4$ into
Eq.~(\ref{fid-ab})
yields
\begin{equation}
F_g=\frac{1}{2}\left[2-(1-S^4)(\abs{a_1}^2+\abs{b_0}^2)-
      (\abs{c_0}^2+\abs{c_1}^2) \right].
\label{fid-abc}
\end{equation}
We can now use the method of Lagrange multipliers
for the remaining two constraints, which gives the
following system of equations
\begin{eqnarray}
\frac{\partial F_g}{\partial |a_0|} +\sum_{i=1}^2 \lambda_i \cdot
\frac{\partial \varphi_i}{\partial |a_0|} & = & 0 \nonumber \\
\frac{\partial F_g}{\partial |b_0|} +\sum_{i=1}^2 \lambda_i \cdot
\frac{\partial \varphi_i}{\partial |b_0|} & = & 0 \nonumber \\
& & \nonumber \\
& ...  & \nonumber
\end{eqnarray}
Let us concentrate on the equations where we differentiate with respect
to the parameters $c_0,c_1$ and
$\langle C_0\ket{C_1}$. Without loss of generality,
we can consider $c_0$ and $\langle C_0\ket{C_1}$ real, while $c_1$ must
be
taken in general complex ($c_1=|c_1| e^{i\delta}$).
The corresponding equations (obtained by differentiating with respect
to $c_0$, $|c_1|$ and $\langle C_0\ket{C_1}$ respectively) give
\begin{eqnarray}
& &-c_0+\lambda_1 \mbox{Re}[|c_1|e^{i\delta}\langle C_0\ket{C_1}]+
\lambda_2 \mbox{Im}[|c_1|e^{i\delta}\langle C_0\ket{C_1}]=0
\label{lm1}\\
& &-\abs{c_1}+\lambda_1 \mbox{Re}[c_0 e^{i\delta}\langle C_0\ket{C_1}]+
\lambda_2 \mbox{Im}[c_0e^{i\delta}\langle C_0\ket{C_1}]=0
\label{lm2}\\
& &\lambda_1 \mbox{Re}[c_0 |c_1|e^{i\delta}]+
\lambda_2 \mbox{Im}[c_0 |c_1|e^{i\delta}]=0
\label{lm4}
\end{eqnarray}
After multiplying Eqs. (\ref{lm1}) by $c_0$, (\ref{lm2}) by $\abs{c_1}$
and (\ref{lm4}) by $\langle C_0\ket{C_1}$ and inserting the last
equation
into the other two
we find $c_0=|c_1|=0$.
We can therefore conclude that $\ket{\alpha}$ and
$\ket{\beta}$ lie in ${\cal H}_{aa,bb}$.

\section{State-dependent cloners from eavesdropping}
\label{app6}

We take as our starting point for these calculations the development in
Refs.~\cite{fuchs1} and \cite{fuchs2} just at the point where the
eavesdropper's probe is restricted to consist of a single qubit.  I.e.,
we
take $\sin\lambda=0$ in those references.  This leaves a two-parameter
family of unitary interactions to be considered.  (Note:  We shall
interchange the symbols $\alpha$ and $\theta$ used in
Refs.~\cite{fuchs1}
and \cite{fuchs2} so as to be consistent with the notation of the
present
paper.)

With this much given, suppose we label the receiver's state for his
qubit after the eavesdropping interaction by
$\rho^{\scriptscriptstyle {\rm A}}_a$ or
$\rho^{\scriptscriptstyle {\rm A}}_b$,
depending upon whether the sender sent state $\ket{a}$ or $\ket{b}$.
Similarly suppose we label the eavesdropper's probe states by
$\rho^{\scriptscriptstyle {\rm E}}_a$ or
$\rho^{\scriptscriptstyle {\rm E}}_b$.  Then, we have from
Eqs.~(86)--(91) and
Eqs.~(98)--(100) of Ref.~\cite{fuchs2} that the matrix elements for
these operators will be:
\begin{eqnarray}
(\hat\rho^{\scriptscriptstyle {\rm E}}_a)_{00}
&=&
\frac{1}{2}\big(1+\cos2\theta\cos2\phi\big)
\label{Roan}
\\
\rule{0mm}{9mm}(\hat\rho^{\scriptscriptstyle {\rm E}}_a)_{01}
&=&
\frac{1}{4}\Big(
(\cos\theta-\sin\theta)^2\sin2(\phi-\alpha)
+(\cos\theta+\sin\theta)^2\sin2(\phi+\alpha)\Big)
\\
\rule{0mm}{9mm}(\hat\rho^{\scriptscriptstyle {\rm E}}_a)_{11}
&=&
\frac{1}{2}\big(1-\cos2\theta\cos2\phi\big)
\;.
\end{eqnarray}
and
\begin{eqnarray}
(\hat\rho^{\scriptscriptstyle {\rm A}}_a)_{00}
&=&
\cos^2\!\theta\cos^2\!\alpha+
\sin^2\!\theta\sin^2\!\alpha
\\
\rule{0mm}{9mm}(\hat\rho^{\scriptscriptstyle {\rm A}}_a)_{01}
&=&
\cos\theta\sin\theta\sin2\phi\cos2\alpha
+\frac{1}{2}\cos2\phi\sin2\alpha
\\
\rule{0mm}{8mm}(\hat\rho^{\scriptscriptstyle {\rm A}}_0)_{11}
&=&
\sin^2\!\theta\cos^2\!\alpha
+\cos^2\!\theta\sin^2\!\alpha\;.
\label{Innish}
\end{eqnarray}
Hermiticity determines the remainder of the matrix elements.
The matrix elements for $\hat\rho^{\scriptscriptstyle {\rm E}}_b$
and $\hat\rho^{\scriptscriptstyle {\rm A}}_b$ are given by the
same expressions except with $\cos\theta$ and $\sin\theta$
interchanged.  With this interaction, the fidelity between the
sender's and receiver's states---i.e., $1-D$ in Eq.~(33) of
Ref.~\cite{fuchs1} and $1-D(U)$ in Eq.~(101) of
Ref.~\cite{fuchs2}---is given by
\begin{eqnarray}
F(\alpha,\phi)
&\equiv&
\bra{a}\rho^{\scriptscriptstyle {\rm A}}_a\ket{a}
=\bra{b}\rho^{\scriptscriptstyle {\rm A}}_b\ket{b}
\\
&=&
\cos^2\!\alpha\,+\,\frac{1}{2}S\cos2\phi
\sin2\alpha
\,-\,\frac{1}{2}S^2\big(1-\sin2\phi\big)\cos\!2\alpha\;.
\label{BoogieShoes}
\end{eqnarray}

Now, it is shown in Refs.~\cite{fuchs1} and \cite{fuchs2} that if this
interaction is to be one for optimizing the tradeoff between the
eavesdropper's information and the fidelity between the sender's and
receiver's quantum states, then $\alpha$ and $\phi$ must satisfy the
relation
\begin{equation}
\tan2\alpha=\frac{S\cos2\phi}{1-S^2(1-\sin2\phi)}\;.
\label{JoyLilly}
\end{equation}
(See Eq.~(52) in Ref.~\cite{fuchs1} and Eq.~(108) in
Ref.~\cite{fuchs2}.)
On the other hand, in order for the optimal eavesdropping solution to
also
be a quantum cloner, it must be the case that
$\hat\rho^{\scriptscriptstyle {\rm E}}_a=
\hat\rho^{\scriptscriptstyle {\rm A}}_a$
and
$\hat\rho^{\scriptscriptstyle {\rm E}}_b=
\hat\rho^{\scriptscriptstyle {\rm A}}_b$.  A little algebra applied to
Eqs.~(\ref{Roan})--(\ref{Innish}) shows that this can occur only when
$\cos2\phi=\cos2\alpha$.  Hence, if there is not to be an inconsistency
with the constraint given by Eq.~(\ref{JoyLilly}), then it must be the
case that the parameter $x\equiv\sin2\phi$ is such that it satisfies,
\begin{equation}
(S+S^2)x^2+(1-S^2)x-S=0\;.
\end{equation}
Solving this quadratic equation and inserting the result into
Eq.~(\ref{BoogieShoes}) gives the fidelity $F_{l,2}$ of
Eq.~(\ref{Herbert}).

As stated in Section III.B, this discussion can be expanded to produce
a
quantum cloner still better with respect to the ``local'' fidelity
criterion
than the one just found.  It is this: we simply set $\phi=\alpha$ in
the
interaction above and ignore the constraint Eq.~(\ref{JoyLilly}) that
the
interaction lead to optimal eavesdropping.  With this,
Eq.~(\ref{BoogieShoes}) reduces to
\begin{equation}
F(\phi)=\frac{1}{2}+\frac{1}{2}(1+S)\!\left((1-S)\cos2\phi+\frac{1}{2}S
\sin4\phi\right)\;.
\label{BernieBoy}
\end{equation}
This expression is maximized when
\begin{equation}
\sin2\phi=\frac{1}{4S}\Big(-1+S+\sqrt{1-2S+9S^2}\Big)\;.
\label{QueenBee}
\end{equation}
Inserting this particular value for $\sin2\phi$ into
Eq.~(\ref{BernieBoy}) gives the expression $F_{l,3}$ reported in
Eq.~(\ref{Yehuda}).
\newpage

\newpage

\end{document}